\documentclass[aps,prl,twocolumn,superscriptaddress]{revtex4-2}
\usepackage{amssymb}
\usepackage{booktabs}
\usepackage{physics}
\usepackage{graphicx}
\usepackage{amsmath}
\usepackage{amsthm}
\usepackage{amsfonts}
\usepackage[T1]{fontenc}
\usepackage{tikz}
\usepackage{array}
\usepackage{multirow}
\usepackage{color}
\usepackage{esint}
\usepackage{bm}
\usepackage{color}
\definecolor{LinkColor}{rgb}{0.75,0.0,0.2}
\usepackage{bbm}
\usepackage{hyperref}
\hypersetup{
	pdfauthor={good guys},
	pdftitle={good title},
	colorlinks=true,
	citecolor=LinkColor,
	linkcolor=LinkColor,
	urlcolor=LinkColor,
}
\usepackage{babel}
\usepackage{titlesec}

\begin{document}
\title{Hilbert subspace imprint: a new mechanism for non-thermalization}
	
\author{Hui Yu}
\affiliation{Beijing National Laboratory for Condensed Matter Physics and Institute of Physics, Chinese Academy of Sciences, Beijing 100910, China}

\author{Jiangping Hu}
\email{jphu@iphy.ac.cn}

\affiliation{Beijing National Laboratory for Condensed Matter Physics and Institute of Physics, Chinese Academy of Sciences, Beijing 100910, China}
\affiliation{Kavli Institute of Theoretical Sciences, University of Chinese Academy of Sciences, Beijing 100190, China}
\affiliation{New Cornerstone Science Laboratory, Beijing 100190, China}

\author{Shi-Xin Zhang}
\email{shixinzhang@iphy.ac.cn}

\affiliation{Beijing National Laboratory for Condensed Matter Physics and Institute of Physics, Chinese Academy of Sciences, Beijing 100910, China}

\date{\today}
 
\begin{abstract}
The search for non-ergodic mechanisms in quantum many-body systems has become a frontier area of research in non-equilibrium physics.
In this Letter, we introduce Hilbert subspace imprint (HSI)—a novel mechanism that enables evasion of thermalization and bridges the gap between quantum many-body scars (QMBS) and Hilbert space fragmentation (HSF). HSI manifests when initial states overlap exclusively with a polynomial scaling (with system size) set of eigenstates. We demonstrate this phenomenon through two distinct approaches: weak symmetry breaking and initial state engineering. In the former case, we observe that ferromagnetic states including those with a single spin-flip display non-thermal behavior under weak $U(1)$ breaking, while antiferromagnetic states thermalize. In contrast, the $Z_{2}$-symmetric model shows thermalization for both ferromagnetic and antiferromagnetic states. In the latter case, we engineer the initial state prepared by shallow quantum circuits that enhance the overlap with the small target subspace. Our results establish HSI as a mechanism equally fundamental to non-thermalization as QMBS and HSF.
\end{abstract}

\maketitle

\textit{Introduction.---} 
Thermalization \cite{deutsch1991quantum,rigol2012thermalization} in quantum many-body systems refers to the process by which a non-equilibrium system evolves toward a state that can be described by equilibrium statistical mechanics. The eigenstate thermalization hypothesis (ETH) \cite{srednicki1994chaos,d2016quantum,deutsch2018eigenstate} provides a framework for understanding thermalization in isolated quantum many-body systems. Specifically, expectation values of observables in eigenstates vary smoothly with energy and match micro-canonical predictions, ensuring that local observables relax to thermal equilibrium values. This explains why most quantum systems equilibrate, despite their underlying unitary dynamics. 

While ETH successfully explains thermalization in generic quantum many-body systems, certain mechanisms lead to violations of ETH with persistent non-thermal behavior. One such pathway is quantum integrability \cite{olshanetsky1983quantum,franchini2017introduction}, where an extensive set of conserved quantities prevents thermalization due to exact solvability and insufficient chaotic dynamics with prominent examples \cite{baxter1972one, lieb1963exact} solvable via the Bethe ansatz \cite{bethe1931theorie}. Another violation of ETH arises in many-body localized (MBL) systems \cite{nandkishore2015many,altman2015universal,abanin2017recent,abanin2019colloquium,alet2018many}, where strong disorder \cite{vznidarivc2008many,pal2010many} or quasiperiodic potentials \cite{iyer2013many,khemani2017two,zhang2018universal} lead to logarithmic entanglement growth \cite{bardarson2012unbounded,deng2017logarithmic,huang2017out,fan2017out,chen2017out,banuls2017dynamics,chen2024subsystem} and emergent local integrals of motion \cite{serbyn2013local,huse2014phenomenology}. 
Another non-thermalization mechanism is prethermalization
\cite{berges2004prethermalization,mori2018thermalization} where systems remain in a long-lived metastable state before eventual thermalization. This occurs either through weak integrability-breaking perturbations \cite{mierzejewski2015identifying,ilievski2015quasilocal,jian2019universal} or under strong periodic driving \cite{kuwahara2016floquet,mori2016rigorous,abanin2017effective}, where emergent quasi-conserved quantities constrain dynamics over exponentially long timescales.

\begin{figure*}
\centering
\includegraphics[width=0.95\textwidth, keepaspectratio]{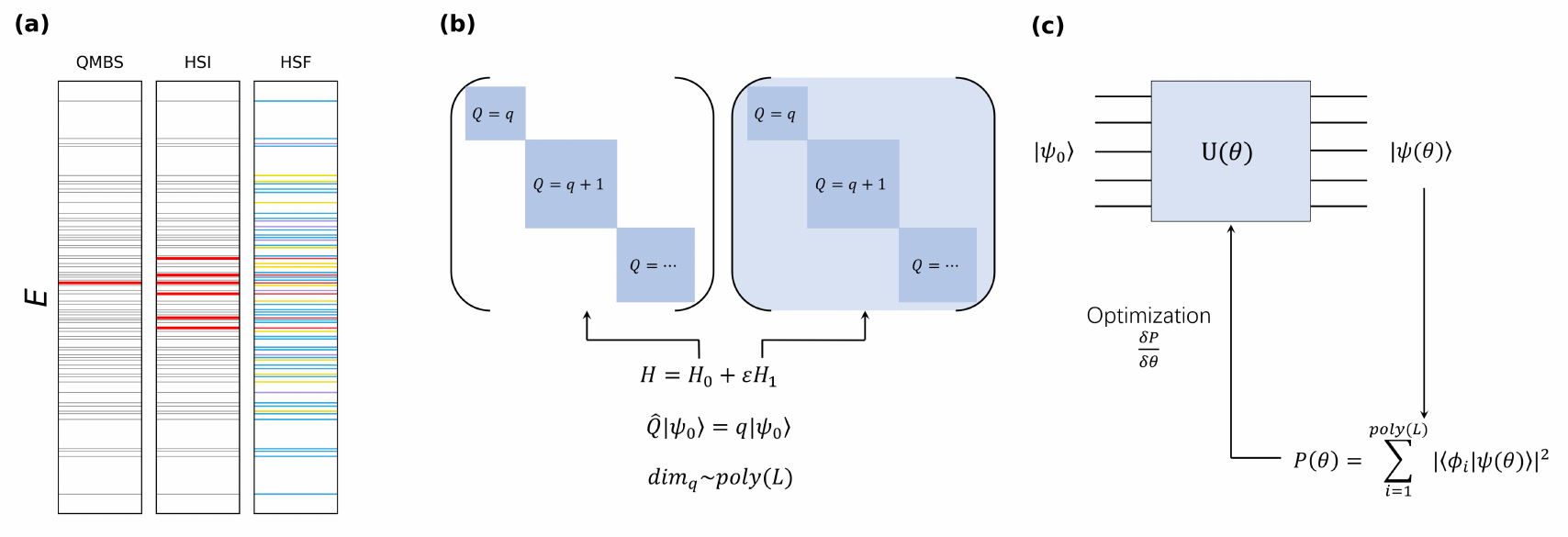}
\caption{Panel (a) displays the energy spectrum for three distinct non-thermal mechanisms. QMBS: The red line marks the scar eigenstate. HSI: Red lines denote a small set of eigenstates restricted to a subspace due to the overlap with certain initial states. HSF: The Hilbert space fractures into disconnected subspaces due to symmetry constraints. Eigenstates of the same color belong to the same subspace except the gray lines for thermal eigenstates. Panel (b) illustrates weak symmetry breaking mechanism leading to HSI. Here, $H_{0}$ is symmetric in terms of symmetry operator $\hat{Q}$, while $H_{1}$ breaks the symmetry, indicated by colors outside the blocks. The symmetry breaking strength is tuned by $\epsilon$. The initial state $|\psi_{0}\rangle$, with symmetry charge $q$, mainly overlaps with $poly(L)$ eigenstates under weak symmetry breaking. Panel (c) demonstrates initial state engineering mechanism for HSI. The initial product state is processed through a shallow variational quantum circuit parameterized by $\theta$. The overlap probability between the resulting state $|\psi(\theta)\rangle$ and target subspace spanned by eigenstates $|\phi_{i}\rangle$ is iteratively optimized until convergence.}
\label{fig:HSI}
\end{figure*}

Beyond these, quantum many-body scars (QMBS) \cite{serbyn2021quantum,chandran2023quantum,moudgalya2022quantum,turner2018weak,turner2018quantum}, which are rare, low-entanglement eigenstates embedded within an otherwise thermal spectrum, provide another route to ergodicity breaking. The paradigmatic example is the PXP model \cite{lesanovsky2012interacting,fendley2004competing}, introduced to describe the Rydberg atom arrays \cite{bernien2017probing} with the nearest-neighbor blockade interactions. Since its discovery, the PXP model has become a cornerstone for understanding QMBS, sparking extensive theoretical and experimental investigations \cite{surace2020lattice,iadecola2019quantum,lin2019exact,rozon2022constructing,giudici2024unraveling,windt2022squeezing,szoldra2022unsupervised,ljubotina2023superdiffusive,desaules2023prominent,desaules2023weak,kerschbaumer2025quantum,wang2024embedding,zhang2023extracting,ren2021quasisymmetry,moudgalya2020eta,ren2025scarfinder,bull2022tuning,gotta2023asymptotic,yao2022quantum,ren2024quasi,ren2022deformed}. Another distinct mechanism for ergodicity breaking is Hilbert space fragmentation (HSF) \cite{moudgalya2022quantum,sala2020ergodicity,khemani2020localization,scherg2021observing,yang2020hilbert,herviou2021many,hahn2021information,moudgalya2022hilbert,moudgalya2022thermalization}, where the Hilbert space dynamically fractures into exponentially many disconnected subsectors known as Krylov subspaces. HSF has been extensively studied in systems with dipole-moment conservation \cite{khemani2020localization,sala2020ergodicity,pai2019localization} as well as other systems with constraints~\cite{de2019dynamics, langlett2021hilbert,richter2022anomalous,zadnik2021folded,read2007enlarged}. Beyond ergodicity breaking, HSF provides a rich platform for novel phenomena including Krylov-restricted thermalization \cite{langlett2021hilbert,schulz2019stark} and subdiffusive dynamics \cite{feldmeier2020anomalous,morningstar2020kinetically,iaconis2019anomalous,iaconis2021multipole}. 

While previous studies have extensively explored non-thermal dynamics through established mechanisms, a fundamental question remains: do alternative pathways exist to suppress thermalization? 
In this Letter, we introduce Hilbert subspace imprint (HSI) as a novel mechanism for stabilizing non-thermal dynamics. HSI occurs when a short-range entangled initial state has significant overlap with only a polynomially small subset of the system's eigenstates. This framework naturally interpolates between QMBS and HSF. On one hand, it generalizes QMBS, which correspond to the special case where the initial state overlaps with a few eigenstates of low entanglement. On the other hand, it describes a form of initial-state-dependent partial fragmentation, distinct from HSF which requires a complete decomposition of the Hilbert space due to symmetry constraints.  Crucially, the eigenstates comprising the HSI subspace are not required to have low entanglement or a specific structure like equal energy spacing, making HSI a more broad and versatile mechanism \footnote{While also initiated from a product state, the non-thermalization from weak Hilbert space fragmentation \cite{yang2025constructingquantummanybodyscars} is distinct. It relies on confinement to a Krylov subspace spanned by product states. In contrast, HSI can describe confinement to a subspace of energy eigenstates, which are typically highly entangled and not product states themselves, representing a more general mechanism.}. A schematic comparison of these mechanisms is presented in Fig.~\ref{fig:HSI}(a).
Crucially, HSI exhibits initial-state-dependent partial fragmentation, where only specific subspaces expanded by several eigenstates remain dynamically closed while others thermalize. How can HSI be naturally realized and how do different initial states respond to HSI?
Addressing these questions will not only deepen our understanding of non-thermal phenomena in quantum many-body systems but also pave the way for harnessing HSI to engineer long-lived coherent states in experiments.

\textit{Setup.---}In this Letter, we demonstrate an approximate version of HSI in which an initial state significantly overlaps with $O(L)$ eigenstates with more than $80\%$ probability weight. The remaining contribution, arising from states outside this subspace, does not grow with time (see Supplementary Material (SM) for derivation). We establish two distinct realization mechanisms for approximate HSI, as illustrated in Fig.~\ref{fig:HSI}(b) and (c): (1) Weak symmetry breaking.  Consider an initial product state $|\psi_{0}\rangle$ that lives exclusively in one charge sector of dimension $poly(L)$ of a symmetric Hamiltonian $H_{0}$, when a weak symmetry-breaking perturbation $H_{1}$ is introduced, the eigenstates of the full Hamiltonian $H=H_{0}+H_{1}$ remain adiabatically connected to those of $H_{0}$. Consequently, $|\psi_{0}\rangle$ retains significant overlap with only $poly(L)$ eigenstates of $H$, leading to desired non-ergodic dynamics. (2) Initial state engineering. The initial state can go beyond the simple product state which is prepared by a shallow-depth parameterized quantum circuit with tunable parameters $\theta$. By optimizing $\theta$, we can obtain the initial state $|\psi(\theta)\rangle$ accessible in experiments of the largest overlap with the target polynomial-sized Hilbert subspace. 

To quantify HSI, we establish a criterion based on the system size scaling of $N$-the minimum number of eigenstates $\{ |E_{i}\rangle \}$ required to capture $80\%$ of the initial state's weight. The system size dependence of $N(L)$ serves as a quantitative indicator: thermalizing states exhibit characteristic exponential scaling $N\sim e^{L}$, while non-thermal states follow at most polynomial growth $N\sim poly(L)$. Moreover, we analyze the time-dependent probability distribution $P_Q(t)$ and entanglement entropy $S_{a}$ for the dynamics. Here, $P_{Q}$ at time $t$ is defined as:
\begin{equation}
P_{Q}(t)=\sum_{q=Q}|\langle \psi(t)|\phi_{q}\rangle|^{2}  
\end{equation}
where $Q$ labels the charge sector, $|\psi(t)\rangle$ is the time-evolved state, and $|\phi_{q}\rangle$ denotes the basis wave function corresponding to charge $q$. The entanglement entropy (EE) $S_{a}$ of subsystem $a$ is given by: $S_{a}=-\text{Tr}(\rho_{a}\log\rho_{a})$, where $\rho_{a}$ is the reduced density matrix obtained by tracing out all degrees of freedom outside $a$. Throughout our analysis, we consider the bipartition where subsystem $a$ comprises exactly half of the total system. Together, these quantities offer complementary signatures of thermalization, allowing us to systematically identify initial states with HSI, distinguishing them from thermalizing dynamics.

\begin{figure}
\centering
\includegraphics[width=0.5\textwidth, keepaspectratio]{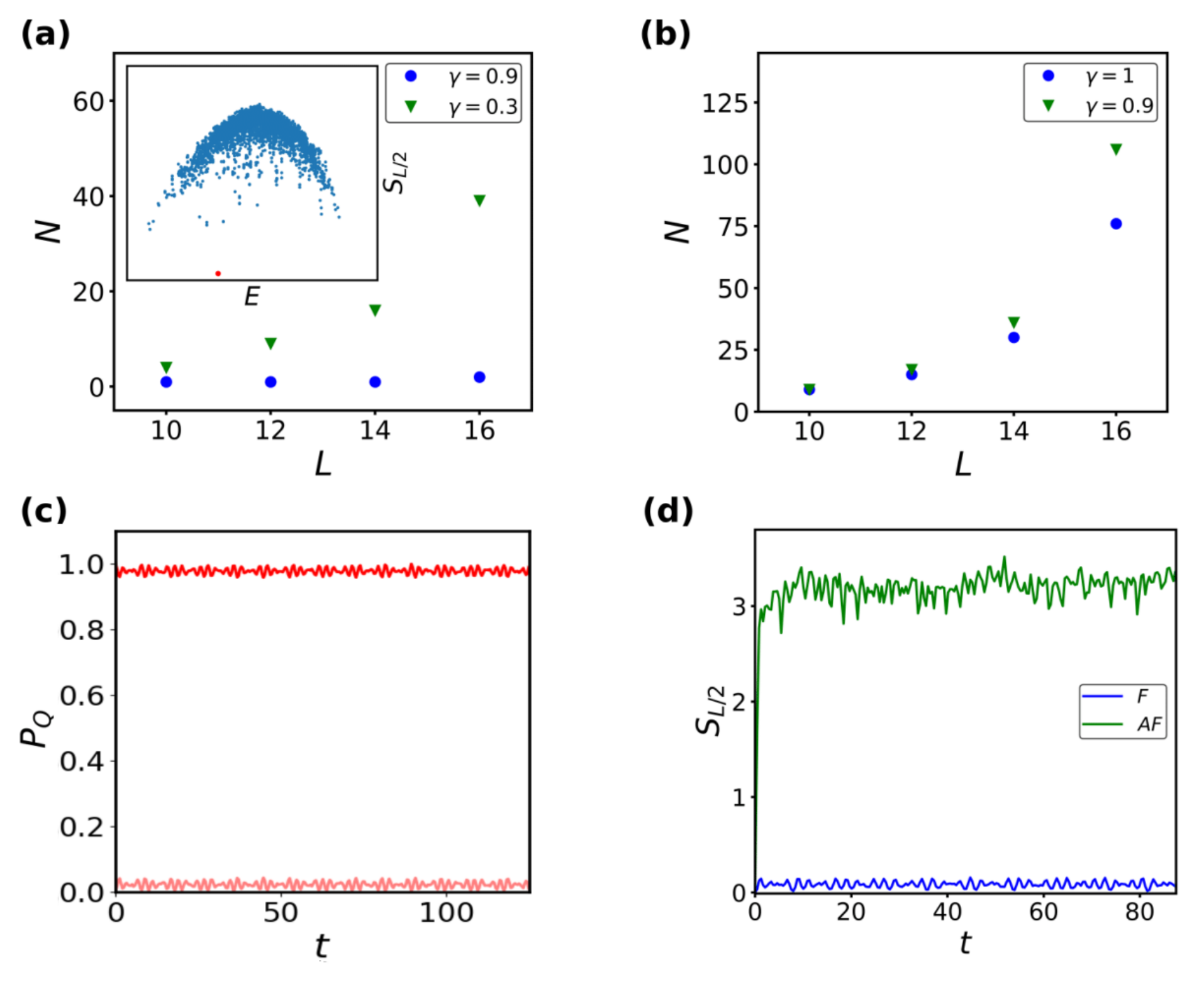}
\caption{Static and dynamical properties for $H_{U1}$ from F and AF initial states (QMBS case). The system-size dependence of $N$ for (a) F and (b) AF initial states at various symmetry-breaking strengths $\gamma$. The inset displays half-chain entanglement entropy $S_{L/2}$ versus energy $E$ at $L=12$, with blue dots marking all eigenstates and the red dot highlighting the scar state ($>80\%$ overlap with initial states). Panel (c) presents the charge distribution dynamics $P_{Q}(t)$ for the F state for $L=12$ at $\gamma=0.9$, where negligible-population sectors ($Q=4,0,-4,-8,-12$) are omitted. Panel (d) compares entanglement entropy dynamics between F and AF states for $L=12$ at $\gamma=0.9$. }
\label{fig:U(1)NNN_FandAF}
\end{figure}

\begin{figure}
\centering
\includegraphics[width=0.5\textwidth, keepaspectratio]{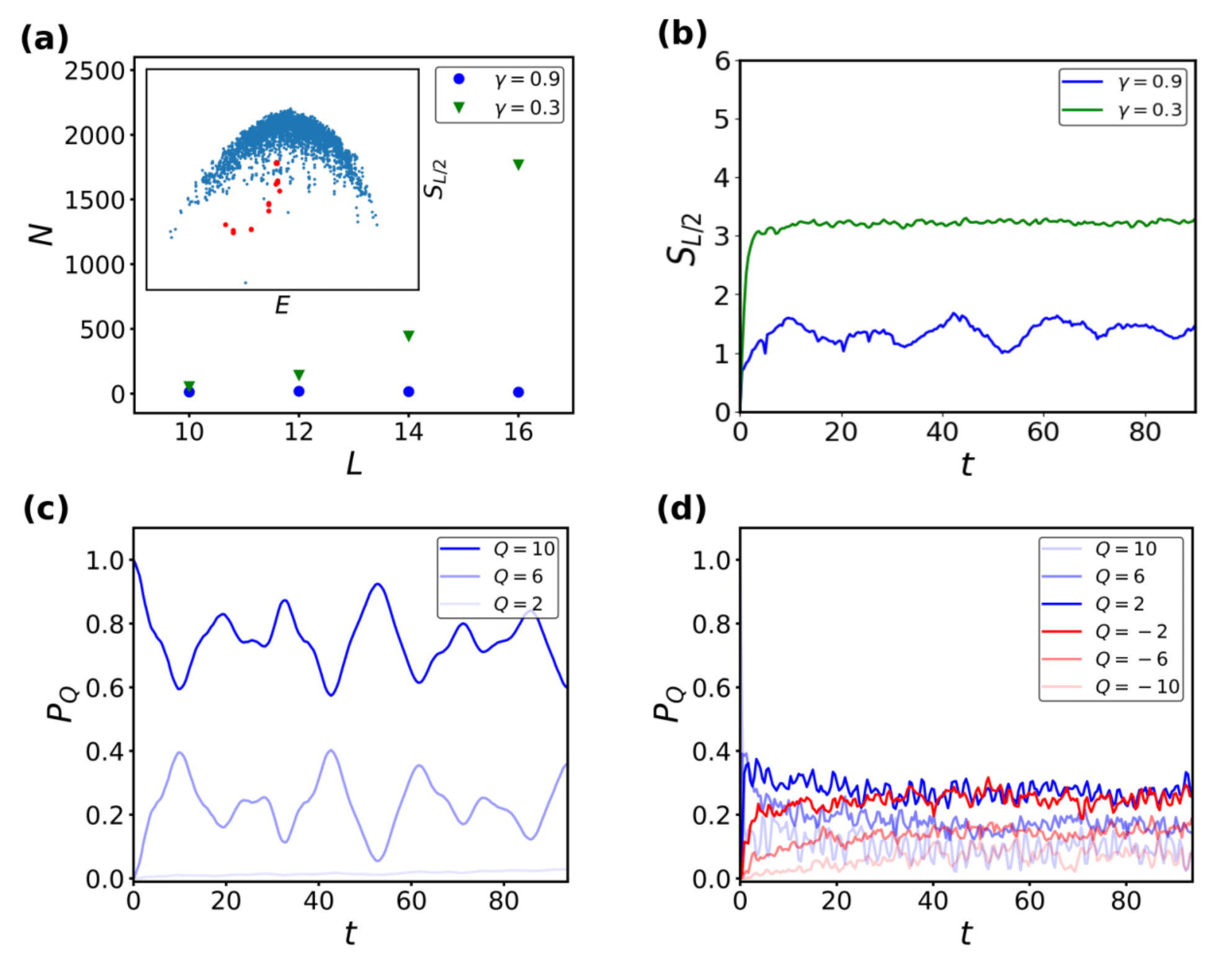}
\caption{Static and dynamical properties for $H_{U1}$ from F state with a single spin flip (HSI case). Panel (a) displays $N(L)$ for the initial spin-flip F state at varying $\gamma$. The inset shows the half-chain entanglement entropy versus energy at $L=12$, with blue dots indicating eigenstates and red dots marking eigenstates contributing to $N$. Panel (b) illustrates the half chain entanglement entropy $S_{L/2}$ dynamics at $\gamma=0.9,0.3$ for $L=12$. Panels (c) and (d) present the charge distribution dynamics $P_{Q}(t)$ at $\gamma=0.9,0.3$ for $L=12$, omitting  $Q=-2,-6,-10$ sectors in (c) due to their negligible populations. All calculations are obtained under $H_{U1}$.}
\label{fig:U(1)NNN_flipspin}
\end{figure}

In our analysis, we examine two initial states: the ferromagnetic (F) state $\vert 000...0 \rangle$ and the antiferromagnetic (AF) state $\vert 0101..1 \rangle$. Additionally, we consider F state with a single spin flip at the chain center. 
To track the evolution of charge sector distributions, we numerically simulate the Hamiltonian dynamics through exact diagonalization: $\vert \psi (t) \rangle=e^{-iHt}\vert \psi_0\rangle$, where $H$ denotes the target Hamiltonian. All simulations are performed using the {\sf TensorCircuit-NG} package \cite{zhang2023tensorcircuit}. 

\textit{Hamiltonian with weak $U(1)$ symmetry breaking.---} We investigate the Hamiltonian with periodic boundary conditions: 
\begin{eqnarray}
H_{U1} = & - \sum_{j=1}^{L} \Big[\sigma_j^x \sigma_{j+1}^x + \gamma \sigma_j^y \sigma_{j+1}^y + \lambda_{1} \sigma_j^z \sigma_{j+1}^z \Big]  \label{eq:Ham1} \\
& -\lambda_{2} \sum_{j=1}^{L} \Big[ \sigma_j^x \sigma_{j+2}^x + \sigma_j^y \sigma_{j+2}^y + \sigma_j^z \sigma_{j+2}^z \Big] 
& +h\sum_{j=1}^{L}\sigma_{j}^z\notag.
\end{eqnarray}
where $\sigma_{j}^{\alpha}$ ($\alpha=x,y,z$) are Pauli matrices acting on site $j$, with $\lambda_{1}=0.2$ and $\lambda_{2}=0.32$ representing the nearest-neighbor and next-nearest-neighbor coupling strengths, respectively. The anisotropy parameter $\gamma$ breaks the system's $U(1)$ symmetry when $\gamma\neq 1$. The conserved $U(1)$ charge is $\hat{Q} = \sum_{i=1}^L \sigma_i^z$. We also include a small longitudinal field $h=10^{-7}$ to lift degeneracies. The NNN interaction is introduced to make the Hamiltonian non-integrable.

We first examine the system-size scaling of $N$ for the F state at $\gamma=0.9$. As shown in Fig.~\ref{fig:U(1)NNN_FandAF}(a), $N$ remains approximately constant as $L$ increases. For $L=12$, only a single eigenstate (with more than $90\%$ overlap with the F state) contributes to $N$. This QMBS state is identified by its anomalous position in the energy spectrum (red dot) and its low half-chain EE, representing a limiting case of HSI. The QMBS nature is manifested in the charge distribution, where the population is predominantly concentrated in the $Q=12$ sector with minimal leakage to $Q=8$, as in Fig.~\ref{fig:U(1)NNN_FandAF}(c). In contrast, the AF state exhibits thermalization behavior. Fig.~\ref{fig:U(1)NNN_FandAF}(b) reveals that $N$ grows exponentially with $L$ even under the $U(1)$-symmetric Hamiltonian. This growth accelerates when the symmetry is weakly broken. The distinct thermal character of the AF state is further reflected in its entanglement dynamics (Fig.~\ref{fig:U(1)NNN_FandAF}(d)), where the saturated EE reaches values significantly higher than the EE of the F state. Notably, when $1-\gamma$ increases to $0.7$ (strong symmetry breaking), the F state also crosses over to thermal behavior. The corresponding time evolution of $P_{Q}$ for this regime is presented in the SM, where the charge distribution spreads across all symmetry-allowed sectors without preferential occupation. 

\begin{figure}
\begin{center}
\includegraphics[width=0.48\textwidth, keepaspectratio]{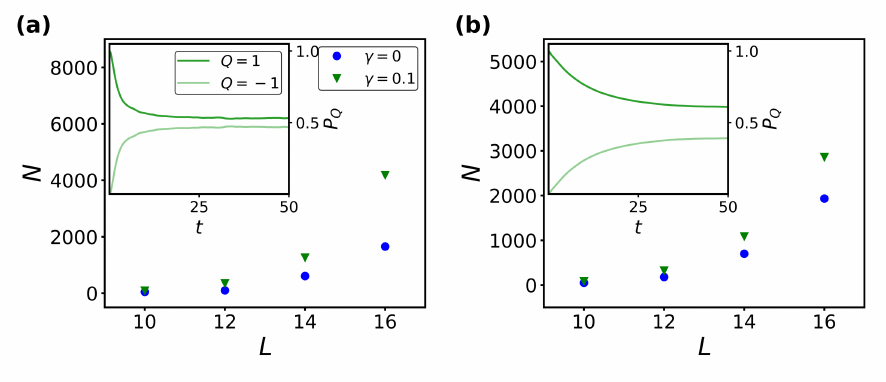}
\caption{Thermalization behavior for $H_{Z2}$ model. $N$ versus system size $L$ for the (a) F and (b) AF initial states with different $\gamma$ values. The insets show the dynamics of charge distributions $P_{Q}(t)$ for $L=12$ at $\gamma=0.1$. All results are averaged over $400$ independent realizations.}
\label{fig:Z2disorder_FandAF}
\end{center}
\end{figure}

Now we proceed to investigate the thermal properties of a spin-flipped F state, initialized with a single spin flip at the $L/2$-th site \footnote{This state can be understood as a localized wavepacket formed by a superposition of the quasi-Nambu-Goldstone modes identified in Ref. \cite{ren2024quasi}. While a single quasi-Nambu-Goldstone mode is a delocalized momentum state, the localized product state in our case serves as an experimentally accessible proxy to probe the non-thermal dynamics.}. Under weak symmetry breaking ($\gamma=0.9$), this state exhibits robust non-thermal behavior, manifested in the polynomial scaling of $N$ with system size shown in Fig.~\ref{fig:U(1)NNN_flipspin}(a). Unlike the F state, the spin-flipped case receives contributions from $O(L)$ eigenstates (red markers), combining both QMBS and non-scar volume-law entangled eigenstates. This multi-state involvement represents the generic HSI scenario for non-thermal dynamics, generalizing the QMBS scenario from the F state case. A clear thermal crossover emerges when increasing the asymmetry to $1-\gamma=0.7$, where $N(L)$ transitions from polynomial to exponential scaling. This thermalization is further corroborated by entanglement dynamics in Fig.~\ref{fig:U(1)NNN_flipspin}(b). In the SM, we further analyze the behavior of the size dependence of steady state EE: while it remains almost constant in the non-thermal regime, it grows linearly with $L$ in the thermal regime, consistent with volume-law scaling. The charge distribution dynamics in Fig.~\ref{fig:U(1)NNN_flipspin}(c) and (d) provide additional distinguishing features. When $\gamma=0.9$, $P_{Q}$ remains sharply peaked in the $Q=10$ and $Q=6$ sectors, confirming the non-thermal character of the state. In contrast, at $\gamma=0.3$ we observe complete thermalization, with $P_{Q}$ distributed across all accessible charge sectors permitted by the symmetry. The enhanced population in $Q=\pm2$ sectors is due to their larger Hilbert space dimensionality. 

We further investigate the dynamics of F and AF states under weak $U(1)$ symmetry breaking in a disordered Hamiltonian. The F state also exhibits persistent non-thermal dynamics, whereas the AF state undergoes rapid thermalization (see SM).

\begin{figure}
\centering
\includegraphics[width=0.48\textwidth, keepaspectratio]{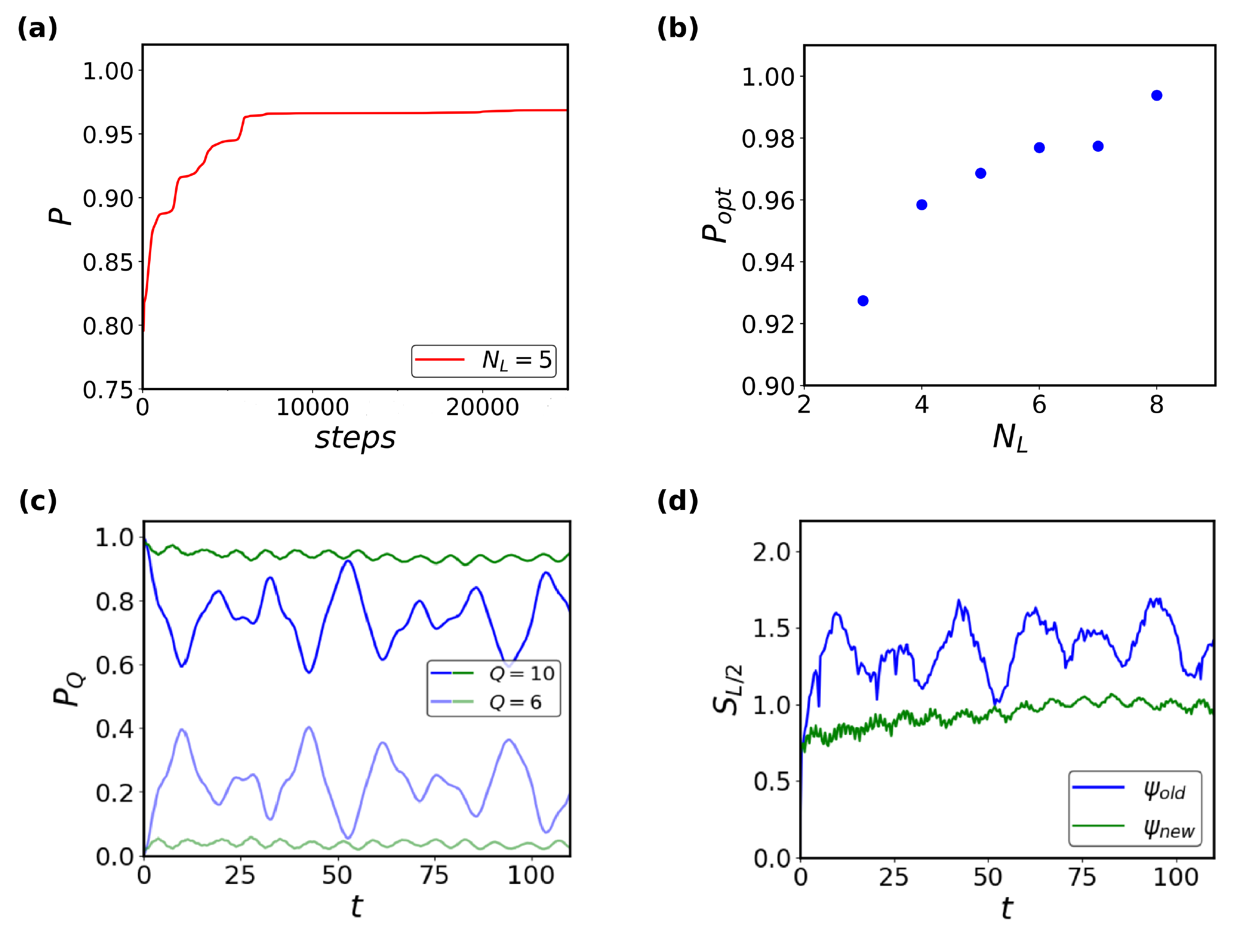}
\caption{Panel (a) tracks the training history of the overlap probability $P$ between the trained state and the target eigenstate subspace for a $5$-composite layers quantum circuit acting on an input state of $11$ spin-up qubits with a single spin-down at the 6th position. Panel (b) shows the optimal overlap probability $P_{opt}$ versus increasing the number of circuit layers. Charge dynamics comparison between pre- (blue) and post-optimization (green) states are presented in panel (c), where negligible-population sectors $Q=2,-2,-6,-10$ are omitted. Panel (d) compares the entanglement entropy dynamics between pre- and post-optimization states. All results in (c) and (d) are obtained for $H_{U1}$.}
\label{fig:Train_flipone}
\end{figure}

\textit{Hamiltonian with weak $Z_{2}$ symmetry breaking.---}
We now study the thermalization properties of F and AF states under a disordered Hamiltonian with explicit breaking of the discrete $Z_{2}$ symmetry. The system is described by:
\begin{eqnarray}
H_{Z2} = & - \sum_{j=1}^{L} \Big[\sigma_j^x \sigma_{j+1}^x  + \Delta_{1} \sigma_j^z \sigma_{j+1}^z +\gamma \sigma_j^x\Big]  \label{eq:Ham2} \\
& - \sum_{j=1}^{L} h_{j}\sigma_j^z \notag.
\label{eq:Ham2}
\end{eqnarray}
where $\Delta_{1}=0.84$ sets the nearest-neighbor coupling strength. The spatial-dependent fields $h_{j}$ are sampled uniformly from $[-W,W]$ with $W=2.4$. The $Z_{2}$ symmetry, is generated by the parity operator $\hat{Q}=\prod_{i=1}^{L}\sigma_{i}^{z}$, is explicitly broken by the $\sigma^{x}$ term with strength $\gamma$, as $\sigma^{x}$ flips spins. All physical quantities are obtained by averaging over multiple disorder realizations. 

Notably, even when the Hamiltonian preserves $Z_{2}$ symmetry ($\gamma=0$), the F initial state exhibits thermalization signatures, evidenced by the exponential scaling of $N(L)$ in Fig.~\ref{fig:Z2disorder_FandAF}(a). This arises from the exponential Hilbert space dimension of the $Q=1$ sector. Introducing $\gamma>0$ enhances thermalization, as the new basis state is a superposition of basis in $Q=1$ and $Q=-1$ sectors. Consequently, the initial states can have nonzero overlap with $Q=-1$ basis states. This fact is further confirmed by computing the time evolution of the sectors $P_{Q=1}$ and $P_{Q=-1}$, as shown in Fig.~\ref{fig:Z2disorder_FandAF}(c). Similar calculations for the AF state reveal analogous thermalization behavior under the same symmetry-breaking conditions, as reflected in Fig.~\ref{fig:Z2disorder_FandAF}(b) and (d).

\textit{Initial state engineering.---}
Recent advances in quantum information processing have enabled high-fidelity state preparation protocols, providing new opportunities to realize and enhance HSI. By implementing a variational quantum circuit \cite{nielsen2010quantum,fisher2023random,bharti2022noisy,cerezo2021variational} (see SM for architecture details), we can engineer the target initial state to better confine it within a small Hilbert subspace spanned by these $O(L)$ eigenstates. The circuit comprises $N_{L}$ layers, each comprising three layers of single-qubit gates followed by one layer of two-qubit gates. We initialize the system in a F state with a single spin flip, as previously described, and train the circuit until the overlap probability with the target eigenstates converges. As revealed in Fig.~\ref{fig:Train_flipone}(a) and (b), the trained state with $N_{L}=5$ achieves $97\%$ overlap weight in the desired subspace. Further increasing $N_{L}$ to $7$ or higher optimizes this overlap to near unity, producing a state with even stronger non-thermal characteristics and nearly perfect HSI. This is further illustrated in Fig.~\ref{fig:Train_flipone}(c) and (d) through both charge probability distribution and EE dynamics. Unlike the untrained state in Fig.~\ref{fig:U(1)NNN_flipspin}(c), the trained state concentrates nearly $100\%$ of its weight in the $Q=10$ sector while displaying lower EE.

\textit{Conclusions and discussions.---} In this Letter, we establish HSI as a fundamental mechanism for non-thermalization, bridging QMBS and HSF. HSI emerges when initial states selectively overlap with a polynomially small subset of eigenstates, and we demonstrate it through two complementary approaches: First, by weakly breaking symmetries, we show that HSI depends on initial states and the symmetry group’s structure. Second, we introduce a constructive, state-centric approach, designing a variational quantum circuit that engineers an initial state to maximize its overlap with a target non-thermal subspace. Both the HSI phenomenon and our proposed realization mechanisms are readily verifiable on current quantum hardware, opening a new avenue for controlling many-body dynamics.

This work opens several avenues for further investigation. A natural extension involves generalizing the framework of weak symmetry breaking to non-Abelian groups, such as 
$SU(2)$. Furthermore, exploring the relationship between HSI and the quantum Mpemba effect~\cite{ares2023entanglement,ares2025quantum,yu2025symmetry,liu2024symmetry,liu2024quantum,chang2024imaginary} promises to yield fundamental insights into the interplay of thermalization and symmetry restoration. Finally, moving beyond the established paradigms of weak symmetry breaking and initial state engineering, an investigation into alternative mechanisms could reveal entirely new strategies for inducing and controlling HSI. 

\textit{Acknowledgement.---}  We acknowledge helpful discussions with Shuo Liu and Jie Ren. We acknowledge the support by the Ministry of Science and Technology  (Grant No. 2022YFA1403900), the National Natural Science Foundation of China (Grant No. NSFC-12494594) and the New Cornerstone Investigator Program. HY is also supported by the International Young Scientist Fellowship of Institute of Physics Chinese Academy of Sciences (No.202407). SXZ is also supported by a start-up grant at IOP-CAS. 

\textit{Data availability.}  Numerical data for this manuscript are publicly accessible in Ref. \cite{data-available}.

\bibliography{reference}

\end{document}


\renewcommand{\thefigure}{S\arabic{figure}}
\setcounter{figure}{0}
\renewcommand{\theequation}{S\arabic{equation}}
\setcounter{equation}{0}
\renewcommand{\thesection}{\Roman{section}}
\setcounter{section}{0}
\setcounter{secnumdepth}{4}
\begin{center}
\textbf{\large Supplemental Material for ``Hilbert subspace imprint: a new mechanism for non-thermalization''}
\end{center}

\author{Hui Yu}
\affiliation{Beijing National Laboratory for Condensed Matter Physics and Institute of Physics, Chinese Academy of Sciences, Beijing 100910, China}

\author{Jiangping Hu}
\email{jphu@iphy.ac.cn}

\affiliation{Beijing National Laboratory for Condensed Matter Physics and Institute of Physics, Chinese Academy of Sciences, Beijing 100910, China}
\affiliation{Kavli Institute of Theoretical Sciences, University of Chinese Academy of Sciences, Beijing 100190, China}
\affiliation{New Cornerstone Science Laboratory, Beijing 100190, China}

\author{Shi-Xin Zhang}
\email{shixinzhang@iphy.ac.cn}

\affiliation{Beijing National Laboratory for Condensed Matter Physics and Institute of Physics, Chinese Academy of Sciences, Beijing 100910, China}
\maketitle

\tableofcontents
\section{More numerical results for $U(1)$-breaking Hamiltonian with NNN terms}

\subsection{The time evolution of charge probability distribution for the antiferromagnetic state at $\gamma=0.9$}

\begin{figure}[ht]
\centering
\includegraphics[width=0.4\textwidth, keepaspectratio]{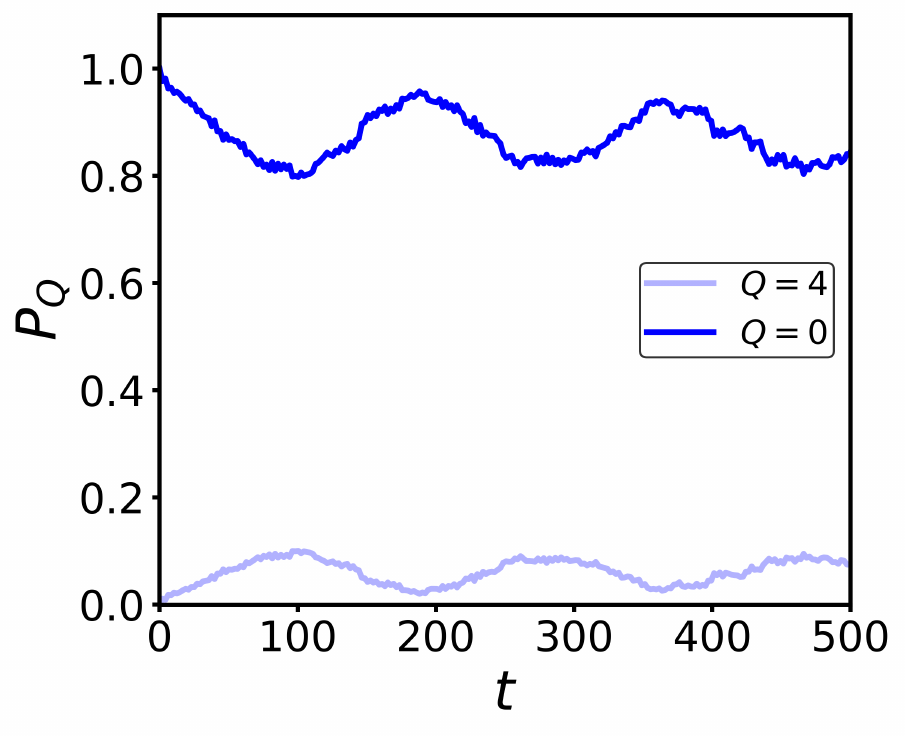}
\caption{The time evolution of charge distributions $P_{Q}(t)$ for AF initial state for $L=12$. $Q=12, 8, -8, -12$ sectors are omitted as they contain negligible population. Note that $P_{Q=-4}$ (not shown) equals $P_{Q=4}$ by symmetry.}
\label{fig:U1NNN_AF_g9}
\end{figure}

Fig.~\ref{fig:U1NNN_AF_g9} displays the time evolution of the charge probability distribution $P_{Q}$ for the antiferromagnetic (AF) state at $\gamma=0.9$. Despite the distribution being confined to only three sectors $Q=0$ and $Q=\pm4$, the AF state exhibits clear thermalization behavior, a consequence of the exponentially large Hilbert space dimensions associated with these sectors, consistent with our findings described in the main text.

\subsection{The time evolution of charge probability distribution for ferromagnetic and antiferromagnetic states at $\gamma=0.3$}
\begin{figure}[ht]
\centering
\includegraphics[width=0.85\textwidth, keepaspectratio]{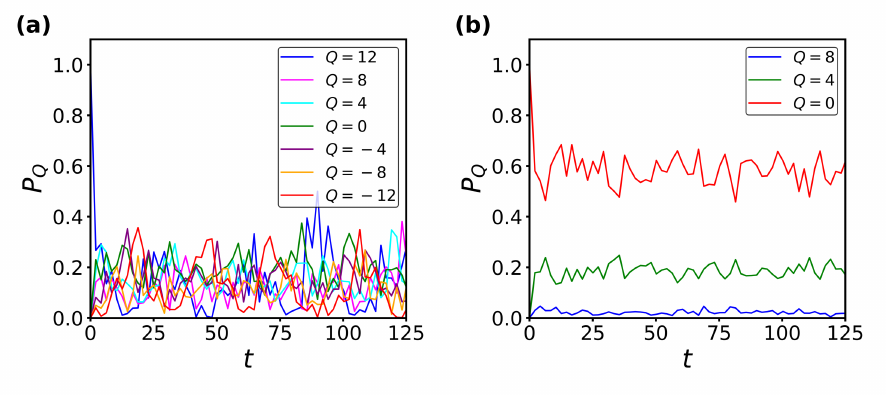}
\caption{The time evolution of charge distributions $P_{Q}(t)$ for (a) F and (b) AF initial state in a system of size $L=12$ with symmetry-breaking strength $\gamma=0.3$. For the AF case, the $Q=12$ and $Q=-12$ sectors are omitted due to negligible population. By symmetry, the $P_{Q=-4}$ and $P_{Q=-8}$ (not shown) distributions are equal to $P_{Q=4}$ and $P_{Q=8}$ respectively.}
\label{fig:U1NNN_FandAF_g3}
\end{figure}

Fig.~\ref{fig:U1NNN_FandAF_g3}(a) demonstrates that the charge probability $P_{Q}$ spreads across all accessible sectors for the ferromagnetic (F) state. indicating the onset of thermalization as symmetry-breaking strength $1-\gamma$ increases. Meanwhile, Fig.~\ref{fig:U1NNN_FandAF_g3}(b) reveals enhanced thermalization in the AF case, evidenced by a more pronounced shift of spectral weight to the $Q=\pm4$ and $\pm8$ sectors compared to Fig.~\ref{fig:U1NNN_AF_g9}.

\subsection{Entanglement entropy scaling in singly spin-flipped ferromagnetic states}
\begin{figure}[ht]
\centering
\includegraphics[width=0.4\textwidth, keepaspectratio]{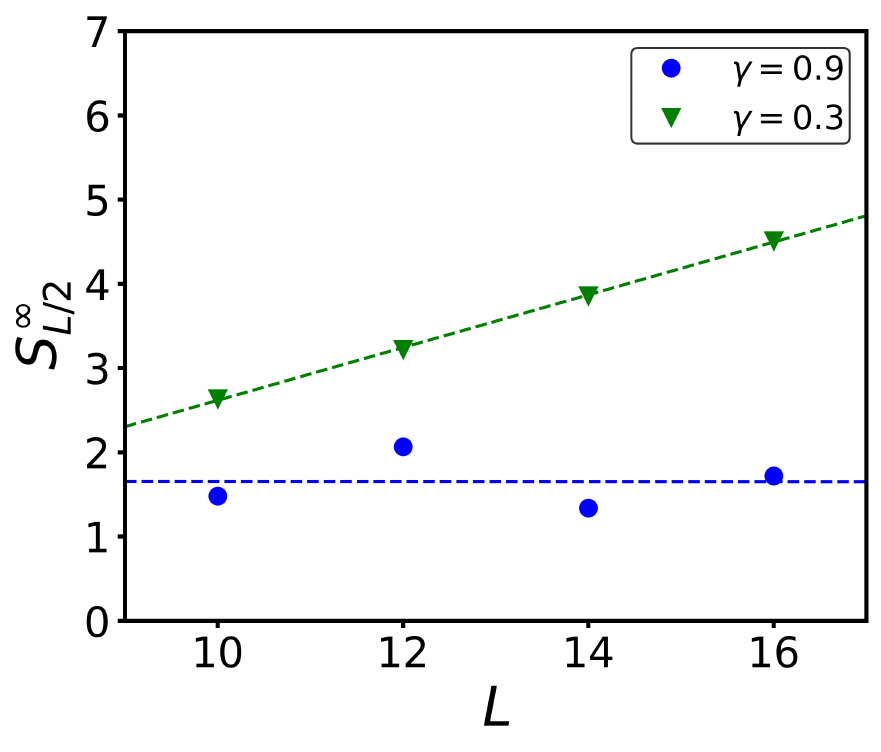}
\caption{The steady-state half-chain entanglement entropy $S_{L/2}^{\infty}$ is plotted as a function of system size $L$ for symmetry breaking strengths $\gamma=0.9$ and $0.3$. Numerical simulation results are shown as data points, while dashed lines indicate the corresponding linear fits to the data.}
\label{fig:U1NNN_EEvsL_flipone}
\end{figure}

Fig.~\ref{fig:U1NNN_EEvsL_flipone} shows the system-size scaling of late-time entanglement entropy for a F state with a single spin flip at the $L/2$-th site under different $\gamma$. At $\gamma=0.9$ (weakly symmetry breaking), the EE remains approximately constant with increasing system size $L$, demonstrating the characteristic area-law scaling of non-thermal states. In contrast, for stronger symmetry breaking $\gamma=0.3$, the EE exhibits volume law linear scaling with system size, a clear signature of thermalization.

\vspace{25cm}

\section{More numerical results for disordered $U(1)$-breaking Hamiltonian}
We now analyze the thermalization properties of F and AF states under a disordered Hamiltonian with weakly broken $U(1)$-symmetry and periodic boundary conditions:
\begin{eqnarray}
H = & - \sum_{j=1}^{L} \Big[\sigma_j^x \sigma_{j+1}^x + \gamma \sigma_j^y \sigma_{j+1}^y + \mu \sigma_j^z \sigma_{j+1}^z \Big]  \label{eq:Ham2} \\
& - \sum_{j=1}^{L} h_{j}\sigma_j^z \notag.
\label{eq:Ham2}
\end{eqnarray}
where $\sigma_{j}^{\alpha}$ ($\alpha=x,y,z$) are Pauli matrices acting on site $j$, with $\mu=0.8$ sets the nearest-neighbor coupling strengths. Disorder is introduced via random fields $h_{j}$ along the $z$-axis, sampled uniformly from $[-W,W]$ with $W=4.8$. The parameter $\gamma$ controls the $U(1)$ symmetry breaking strength. Notably, for $\gamma=1, \mu=0$, this Hamiltonian reduces to the one dimensional Aubry-Andr\'{e} model \cite{lee2017many,chandran2017localization,nag2017many,lev2017transport,vznidarivc2018interaction,crowley2018quasiperiodic}, which exhibits a well-known localization transition. All physical quantities are obtained by averaging over multiple disorder realizations.

\subsection{Thermalization properties for the ferromagnetic state at different $\gamma$}
\begin{figure}[ht]
\centering
\includegraphics[width=0.95\textwidth, keepaspectratio]{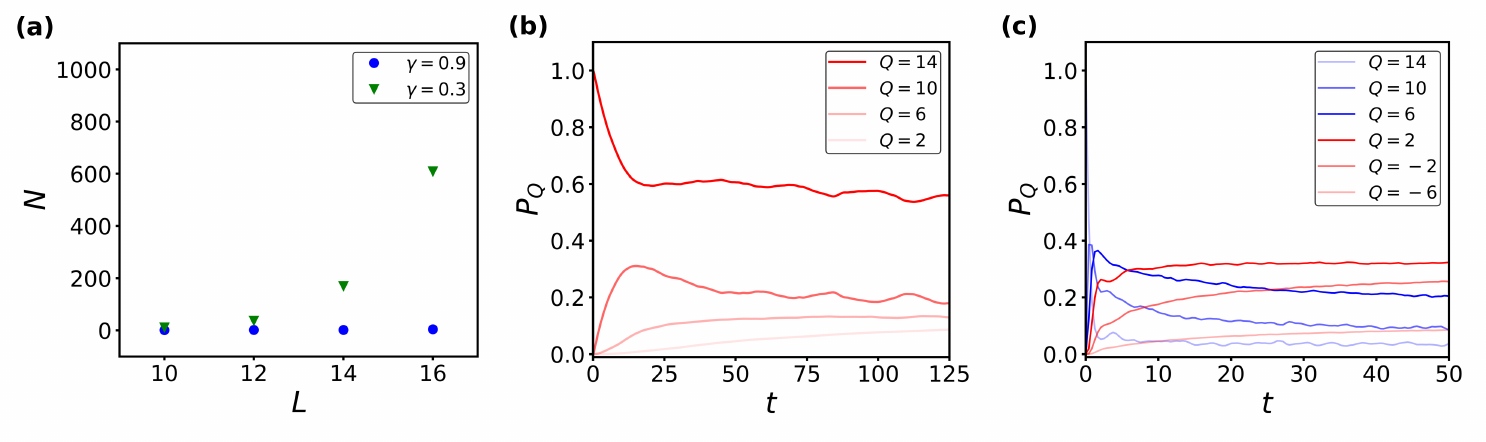}
\caption{Panel (a) displays the scaling of $N$ with system size for F initial state at varying $\gamma$. Panels (b) and (c) show the time evolution of charge distributions $P_{Q}(t)$ initialized in F states for $L=14$ at $\gamma=0.9$ and $\gamma=0.3$. Sectors $Q=12,8,-8,-12$ and $Q=12,-12$ are omitted in panels (b) and (c) due to negligible weight. By symmetry, the $P_{Q}=-4$ and $P_{Q}=-8$ (not shown) distribution are equal to $P_{Q}=4$ and $P_{Q}=8$ respectively. All results are averaged over $400$ independent realizations.}
\label{fig:U1disorder_F}
\end{figure}

Fig.~\ref{fig:U1disorder_F}(a) demonstrates that for $\gamma=0.9$, the F state exhibits clear non-thermal behavior, as characterized by the nearly system-size-independent of $N$. The time evolution of the charge probability distribution in Fig.~\ref{fig:U1disorder_F}(b) further confirms the non-thermal nature, showing approximately $60\%$ of the population remains confined to the Q=14 sector of dimension $1$, with an additional $20\%$ in the Q=10 sector. Upon increasing the symmetry-breaking strength $1-\gamma$ to $0.7$, the F state begins to thermalize, as supported by the exponential growth of $N$ with $L$ and a significantly broader charge distribution shown in Fig.~\ref{fig:U1disorder_F}(c). 

\subsection{Thermalization properties for the antiferromagnetic state at different $\gamma$}
\begin{figure}[ht]
\centering
\includegraphics[width=0.95\textwidth, keepaspectratio]{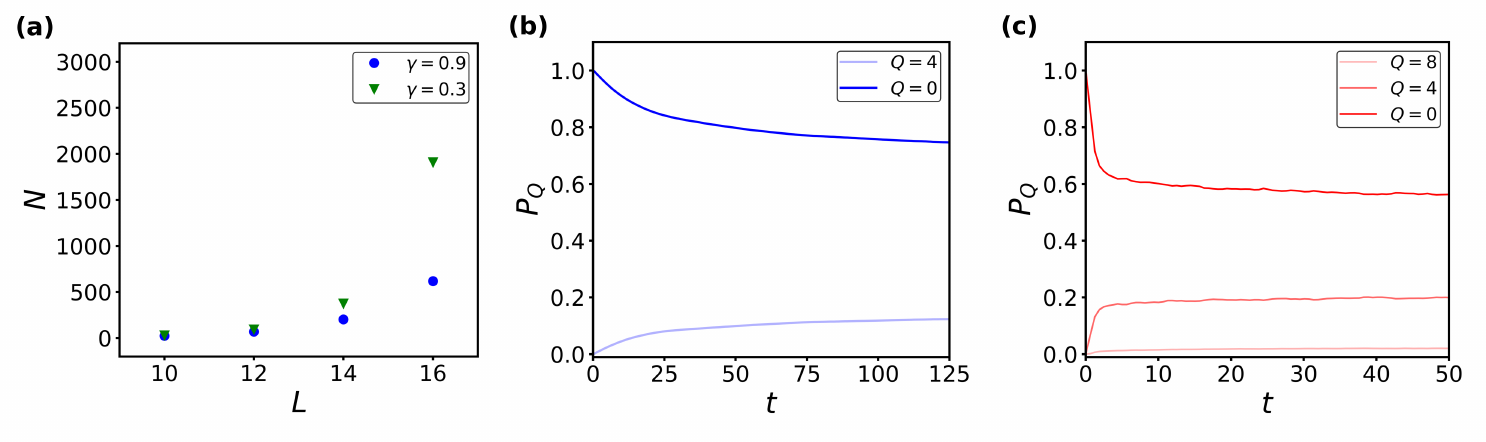}
\caption{Panel (a) displays the scaling of $N$ with system size for AF initial state at varying $\gamma$. Panels (b) and (c) show the time evolution of charge distributions $P_{Q}(t)$ initialized in AF states for $L=14$ at $\gamma=0.9$ and $\gamma=0.3$, respectively. Sectors $Q=-2,-6,-10,-14$ and $Q=-10,-14$ are omitted in panels (b) and (c) due to negligible weight. By symmetry, the $P_{Q=-4}$ and $P_{Q=-8}$ (not shown) distributions are equal to $P_{Q=4}$ and $P_{Q=8}$ respectively. All results are averaged over $400$ independent realizations.}
\label{fig:U1disorder_AF}
\end{figure}

In contrast to the F state, AF initial state shows thermalization under weak symmetry breaking, as confirmed by the exponential growth of $N$ with $L$ in
Fig.~\ref{fig:U1disorder_AF}(a). This growth becomes more pronounced with stronger symmetry breaking. Despite thermalization, Fig.~\ref{fig:U1disorder_AF}(b) confirms the charge distribution remains dominated by the $Q=0$ and $Q=\pm4$ sectors, a consequence of their large Hilbert space dimensions. As the symmetry breaking becomes more pronounced, Fig.~\ref{fig:U1disorder_AF}(c) displays a systematic charge redistribution of spectral weight toward the $Q=\pm 4$ and $Q=\pm 8$ sectors.

\section{The variational quantum circuit ansatz}
\begin{figure}[ht]
\centering
\includegraphics[width=0.75\textwidth, keepaspectratio]{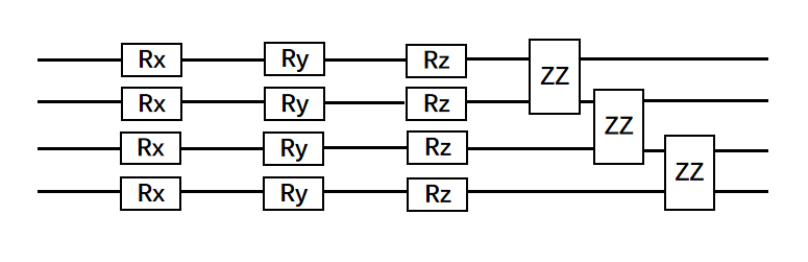}
\caption{Schematic illustration of a trained 4-qubit quantum circuit with one composite layer. Each composite layer comprises: (1) single-qubit rotation gates ($R_{x}$, $R_{y}$, $R_{z}$) applied to all qubits, and (2) two-qubit $ZZ$ coupling gates between neighboring qubits. The circuit is optimized to enhance non-thermal behavior of the target state.}
\label{fig:VQE_circuit}
\end{figure}

Fig.~\ref{fig:VQE_circuit} presents one layer of the variational quantum circuit designed to optimize the initial state for enhanced non-thermal properties. The architecture comprises $N_{L}$ composite layers, each containing three layers of parametrized single-qubit rotation gates $R_{\alpha}(\theta)=e^{-i\sigma_{\alpha}\theta/2}$ ($\alpha=x,y,z$) applied to every qubit, followed by a layer of two qubit $ZZ(\theta)=e^{-i\sigma^{z}_{i}\sigma^{z}_{i+1}\theta}$ gates acting on adjacent qubit pairs ($i$,$i+1$). Each rotation angle $\theta$ is independently optimized through variational methods to maximize the overlap probability $P(\theta)=\sum_{i=1}^{poly(L)}|\langle\phi_{i}|\psi(\theta)\rangle|^{2}$, where $|\phi_i\rangle$ is the $i$-th eigenstate in the target subspace and $|\psi(\theta)\rangle$ is the output state.

\section{Analytical derivation of the bounds on the expectation value $\langle\hat{O}\rangle$}
Suppose an initial wavefunction $|\phi_{0}\rangle$ with significant overlap with an eigenstate in a small non-thermal Hilbert subspace, along with a small leakage to state outside this subspace that can thermalize. We aim to determine the bounds for the expectation value of an observable $\hat{O}$. The initial state can be expressed as:
\begin{equation}
|\psi_{0}\rangle=\sqrt{1-\Delta^{2}}|\phi\rangle+\Delta|\phi_{\perp}\rangle
\end{equation}
where $|\phi\rangle$ is the normalized state in the non-thermal subspace, $|\phi_{\perp}\rangle$ is the normalized state outside that space, which ultimately thermalize. $\Delta^{2}$ quantifies the leakage probability. And we omit the irrelevant relative phase. The time evolution of $|\psi_{0}\rangle$ is given by:
\begin{equation}
|\psi(t)\rangle=\sqrt{1-\Delta^{2}}|\phi(t)\rangle+\Delta |\phi_{\perp}(t)\rangle
\end{equation}
with $|\phi(t)\rangle$ and $|\phi_{\perp}(t)\rangle$ still inside and outside the Hilbert subspace, respectively.

Next, we calculate the expectation value of an observable $\hat{O}$ with respect to the state $|\psi(t)\rangle$ as:
\begin{equation}
\langle \hat{O}\rangle=(1-\Delta^{2})O_{non-th}(t)+\Delta^{2}O_{th}+\Delta\sqrt{1-\Delta^{2}}\left[\langle\phi(t)|\hat{O}|\phi_{\perp}(t)\rangle+\langle\phi_{\perp}(t)|\hat{O}|\phi(t)\rangle\right]
\label{expectO}
\end{equation}
where $O_{non-th}(t)=\langle\phi(t)|\hat{O}|\phi(t)\rangle$ is the expectation value of $\hat{O}$ in the non-thermal subspace and $O_{th}=\langle\phi_{\perp}(t)|\hat{O}|\phi_{\perp}(t)\rangle$ is the thermal expectation value at late times according to ETH. To bound the cross terms, we apply the Cauchy-Schwarz inequality, which states that for any state $|\phi(t)\rangle$, $|\phi_{\perp}(t)\rangle$, and an operator $\hat{O}$, we have:
\begin{equation}
|\langle\phi(t)|\hat{O}|\phi_{\perp}(t)\rangle| \leq ||\hat{O}||\cdot||\phi(t)||\cdot||\phi_{\perp}(t)|| 
\label{spectral-ineq}
\end{equation}
Here, $||\cdot||$ denotes the operator and vector norms. Since $|\phi(t)\rangle$ and $|\phi_{\perp}(t)\rangle$ are normalized ($||\phi(t)||=||\phi_{\perp}(t)||=1$), and $\hat{O}$ is bounded with eigenvalues $\lambda_{k}$ satisfying $|\lambda_{k}|\leq1$ (e.g., Pauli string operators), the operator norm $||\hat{O}||$ is bounded by the largest eigenvalue:
\begin{equation}
||\hat{O}|| \leq 1
\end{equation}
Thus, Eq.~\eqref{spectral-ineq} simplifies to: 
\begin{equation}
|\langle\phi(t)|\hat{O}|\phi_{\perp}(t)\rangle| \leq 1
\label{loose_bound}
\end{equation}
Similarly, $|\langle\phi_{\perp}(t)|\hat{O}|\phi(t)\rangle| \leq 1$. By omiting higher orders of $\Delta$ for small leakage, we have
\begin{equation}
O_{non-th}-2\Delta \leq \langle \hat{O}\rangle\leq O_{non-th}+2\Delta,
\end{equation}
namely, non-thermal behavior can be determined via local observable measurements as long as $\Delta$ is below some threshold.

In fact, we can derive a significantly tighter bound by leveraging the ETH. Let us expand $|\phi(t)\rangle$ and $|\phi_{\perp}(t)\rangle$ in the energy eigenbasis:
\begin{equation}
|\phi(t)\rangle = \sum_{n} d_{n} e^{-iE_{n}t} |E_{n}\rangle, \quad 
|\phi_{\perp}(t)\rangle = \sum_{\alpha} c_{\alpha} e^{-iE_{\alpha}t} |E_{\alpha}\rangle
\end{equation}
where ${|E_{n}\rangle}$ are non-thermalizing eigenstates, ${|E_{\alpha}\rangle}$ are thermalizing eigenstates, and $d_{n}$,~$c_{\alpha}$ are their respective coefficients. The cross term then becomes:
\begin{equation}
\langle\phi(t)|\hat{O}|\phi_{\perp}(t)\rangle = \sum_{n,\alpha}d_{n}^{*}c_{\alpha}e^{i(E_{n}-E_{\alpha})t}\langle E_{n}|\hat{O}|E_{\alpha}\rangle
\end{equation}
The phase factors $e^{i(E_{n}-E_{\alpha})t}$ oscillate rapidly at late times, causing dephasing and suppression of the sum. By the ETH, the magnitude of off-diagonal matrix elements $\langle E_{n}|\hat{O}|E_{\alpha}\rangle$ can be expressed as:
\begin{equation}
|\langle E_{n}|\hat{O}|E_{\alpha}\rangle| \sim e^{-S(\bar{E})/2}f(\bar{E},\omega)
\end{equation}
Here $\bar{E}=\frac{E_{n}+E_{\alpha}}{2}$ represents the average energy, and $\omega=E_{n}-E_{\alpha}$ is the energy difference. The term $S(\bar{E})$ corresponds to the thermodynamic entropy evaluated at energy $\bar{E}$, which is proportional to the system size $L$ while $f(\bar{E},\omega)$ is a smooth function of $\bar{E}$ and $\omega$. Consequently, Eq.~\eqref{loose_bound} is 
\begin{equation}
|\langle\phi(t)|\hat{O}|\phi_{\perp}(t)\rangle| \leq \delta
\label{tight_bound}
\end{equation}
where $\delta$ is an exponentially small parameter with respect to the system size due to strong dephasing and exponential small terms from off-diagonal prediction of ETH. Consequently,
\begin{equation}
(1-\Delta^{2})O_{non-th}+\Delta^{2}O_{th}-2\delta\Delta\sqrt{1-\Delta^{2}} \leq \langle \hat{O}\rangle \leq (1-\Delta^{2})O_{non-th}+\Delta^{2}O_{th}+2\delta\Delta\sqrt{1-\Delta^{2}}
\end{equation}
As $\delta$ is much smaller than $\Delta$ for a large system, we omit $\delta\Delta$ term while keep the quadratic $\Delta$ term instead. We thus have:
\begin{equation}
\langle \hat{O}\rangle-O_{th}\sim (1-\Delta^2)(O_{non-th}-O_{th})
\end{equation}
This relation demonstrates that if $O_{th}$ deviates from $O_{non-th}$, we can unambiguously distinguish between thermal and non-thermal behaviors through the observable expectation value.

\bibliography{reference}